\documentclass[aps,prl,twocolumn,superscriptaddress,showpacs]{revtex4}
\usepackage[pdftex]{graphicx}
\usepackage{latexsym,amsmath,verbatim}
\usepackage{color}
\usepackage{xfrac}
\usepackage{hyperref}
\usepackage{epstopdf, epsfig}

\newcommand*\rfrac[2]{{}^{#1}\!/_{#2}}

\newcommand{\be}{\begin{equation}}
\newcommand{\ee}{\end{equation}}

\newcommand{\bea}{\begin{eqnarray}}
\newcommand{\eea}{\end{eqnarray}}

\begin{document}

\title{Damage detection via shortest path network sampling}

\author{Fabio Ciulla}
\affiliation{Laboratory for the Modeling of Biological and Socio-technical Systems, Northeastern University, Boston MA 02115 USA}

\author{Nicola Perra}
\affiliation{Laboratory for the Modeling of Biological and Socio-technical Systems, Northeastern University, Boston MA 02115 USA}

\author{Andrea Baronchelli}
\affiliation{Department of Mathematics, City University London, Northampton Square, London EC1V 0HB, UK}

\author{Alessandro Vespignani}
\affiliation{Laboratory for the Modeling of Biological and Socio-technical Systems, Northeastern University, Boston MA 02115 USA}
\affiliation{Institute for Scientific Interchange (ISI), Torino, Italy}

\date{\today} \widetext

\begin{abstract}
Large networked systems are constantly exposed to local damages and failures that can alter their functionality. The knowledge of the structure of these systems is however often derived through sampling strategies whose effectiveness at damage detection has not been thoroughly investigated so far. Here we study the performance of shortest path sampling for damage detection in large scale networks. We define appropriate metrics to characterize the sampling process before and after the damage, providing statistical estimates for the status of nodes (damaged, not-damaged). The proposed methodology is flexible and allows tuning the trade-off between the accuracy of the damage detection and the number of probes used to sample the network. We test and measure the efficiency  of our approach considering both synthetic and real networks data. Remarkably, in all the systems studied the number of correctly identified damaged nodes exceeds the number false positives allowing to uncover precisely the damage.
\end{abstract}


\maketitle

Real world networks are often the result of a self-organized evolution without central control~\cite{Newmanlibro,Caldarellilibro,Alexlibro}.  The physical Internet and the World Wide Web are two prototypical examples where the interplay of local and non-local evolution mechanisms defines the global structure of the network. In absence of defined blueprints, the only way to characterize the global structure of self-organizing systems is by devising sampling experiments, like \emph{traceroute} probing for the physical Internet~\cite{MaoTAA2003,Luckie2008,Huffaker2002,LuckieSSE2010,VanJacobson,SpringMIT2004} and \emph{web crawlers} for the WWW~\cite{Bing2011,MenczerTWC2004,mirtaheri2013crawlingSurvey}, that try to infer the properties and connectivity patterns of these structures~\cite{caida,dimes,asdata,lucent,rocketfuel,oregon,Bu:2002:NTG:511399.511338}. 
However, sampling processes are limited by time and physical constraints, and in general guarantee  access only to a part of the network.
In this context, the detection of network damage is a difficult task. The lack of information on the exact structure of the system makes it extremely difficult to identify what damage has been actually suffered and discriminate damaged elements from those that have been simply not yet probed by the sampling process.  

Here we study numerically the effectiveness of shortest path sampling strategies for damage detection in large-scale networks.  Shortest path sampling strategies are actually used in Internet probing~\cite{alex-asta,Newmanlibro,PsV04, claffy_nature, Shavitt_2005,DBLP:journals/corr/abs-1209-5074} including failure detection, via traceroute like tools and end-to end measurements. We consider different attack strategies, namely random or connectivity based and introduce a global measure, $M$, that allows to quickly identify damages that induce large variations in the routing patterns of networks. We then propose a statistical method able to classify nodes as damaged or functioning in the case of partial network sampling. In our analysis, we first sample a given network structure via shortest path probes \cite{alex-asta,Newmanlibro,PsV04, claffy_nature, Shavitt_2005,DBLP:journals/corr/abs-1209-5074} obtaining a partial representation of its nodes and connectivity patterns. Then we damage the network by removing nodes according to different strategies and sample again the damaged network supposing not to know the location and magnitude of  the damage. During this sampling, we constantly monitor whether the probability not to have seen a node exceeds the expected value calculated on the basis of the sampling of the undamaged graph. We define a statistical criterion to asses which nodes of the network can be considered damaged, and we test the performance of the method by looking at the number of true and false positives it identifies. 

We perform numerical experiments on synthetic networks, either heterogeneous, generated with the uncorrelated configurational model (UCM)~\cite{UCM}, or homogeneous, obtained through the Erd\"os-R\`eny model (ER)~\cite{Erdos}. Remarkably, the magnitude and location of the damage can be detected with fairly good confidence. The accuracy improves for nodes that play a central role in network's connectivity. Namely, the detection of damaged hubs is more reliable than the one of peripheral nodes. 
As a practical application, we consider the damage detection on the physical Internet at the level of Autonomous System (AS). We use the AS topology provided by the Dimes project~\cite{dimes}. In this case, to simulate realistic damages, we remove nodes according to their geographical position. In doing so we simulate critical events such as large-scale power outages,  deliberate servers switch offs~\cite{Dainotti_2011} or other major localized catastrophic events~\cite{TechRepVirginiaTech2011}.
Interestingly, also in this case our methodology allows to statistically identify  the extent and location of the damage with reasonable accuracy.

The paper is organized as follows: in the section I we present the sampling method used. In section II we introduce a measure that provides a general estimation of the damage extension in sampled networks. In section III we provide a method to infer if a single node is damaged or not using a p-value test. In section IV we validate this method by applying it to the physical Internet network. Finally in section V we present our conclusions and final remarks.\\


\section{Shortest path sampling of undamaged networks}
\label{sampling}

The sampling of networks via shortest paths consists in sending probes from a set of nodes that have been defined as sources toward another set of nodes chosen to be targets.  Each probe travels through the network following a shortest path and records each node and link visited returning a path. This is, to a first approximation, what is executed by Internet mapping projects that use the traceroute tool~\cite{VanJacobson} as probing method.
This methodology infers paths by transmitting a sequence of limited-time-to-live TCP/IP packets from a source node to a specified target on the Internet. The nodes visited along the way send their IP addresses as a response and create the path. The union of all the paths returned by the probes creates the sampled picture of the network.
However, mapping the real physical Internet is complicated and existing approaches have major limitations~\cite{MaoTAA2003}. For example, visited nodes can fail to provide their IP address, and wrong or outdated forwarding routes registries can result in not optimal forwarding route indications.
Our study is inspired by the traceroute tool although we assume that our probes follow shortest paths in the network neglecting the real world limitations aforementioned.

Here we focus on undirected and unweighted networks constituted of $N$ nodes and $M$ edges $\mathcal{G}(N,M)$. We fix a set of sources $\mathbf{S}=(s_1,s_2,\ldots,s_{N_S})$ and a set of target $\mathbf{T}=(t_1,t_2,\ldots,t_{N_T})$ among the $N$ nodes, being $N_S$ and $N_T$ the total number of sources and targets respectively. For each pair of nodes, taken in the two sets, a shortest path probe is sent. After all of the $N_S \times N_T$ probes are executed, all the resulting paths are merged in a sampled network that we denote as $\mathcal{G}^*(N^*,M^*)$. Here and in the rest of the paper the star symbol indicates sampled quantities, so $N^*$ is the number of discovered nodes and $M^*$ the number of discovered links via to the shortest path sampling.

In general, for each source-destination pair we can have two or more equivalent shortest paths. More precisely, there could be different strategies to numerically simulate the shortest path probing:
\begin{itemize}
\item Unique shortest path (USP). The shortest path between a node $i$ and a target $T$ is always the same independently of the source $S$. Each shortest path is selected initially and they will never change.
\item Random shortest path (RSP). Each shortest path is selected every time randomly among the equivalent ones.
\item All shortest path (ASP). All possible, equivalent, shortest path between shortest path are discovered.
\end{itemize}

\begin{figure}
\begin{centering}
\includegraphics[width=0.4\textwidth]{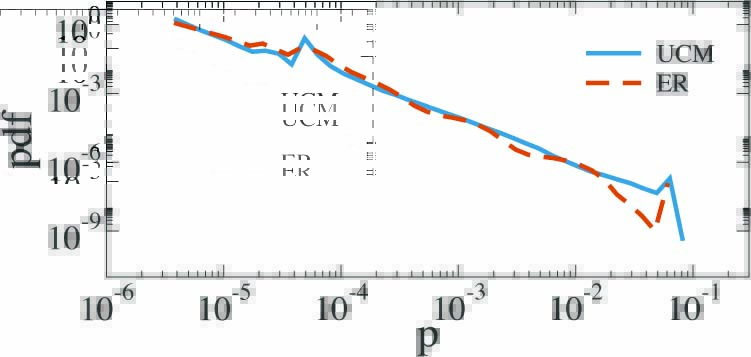}
\par\end{centering}
\caption{\label{probability} {\small Probability distribution function (pdf) of node visit probability for undamaged UCM (blue solid line) and ER (orange dashed line) networks. The two peaks that deviate from the overall heavy-tailed behavior occurs at $p_i = 1/N_S$ and $p_i = 1/N_T$, and represent the  visit probability of sources and target respectively. The curves are the average over $100$ independent simulations.}}
\label{fig:pi}
\end{figure}

\noindent In the following we will use the RSP probing strategy~\cite{alex-asta,crovella,delos,achil-09}.
Both sources and targets are chosen randomly among all the nodes. Inspired by real Internet probing we investigate scenarios in which the order of magnitude of sources is $N_S = \mathcal{O}(10)$, while the order of magnitude of density of targets, $\rho_T$, is $\mathcal{O}(10^{-1})$. 
Along with the raw number of discovered nodes, we also keep track of the visit probability $p_i$ for each visited node $i$, defined as the ratio between the number of shortest path probes passed through the node $i$ and the total number of probes sent $N_S \times N_T$:
\be
p_i=\frac{\sum \limits_{j=1}^{N_S \times N_T} \delta _{i,j}}{N_S N_T},
\label{eq:pi}
\ee
where $\delta_{i,j}$ is equal to $1$ if the node $i$ is seen by the probe $j$.
In the limit in which both $N_S$ and $N_T$ approach $N$, the probability $p$ become the betweenness~\cite{alex-asta}. Instead, in  more realistic cases in which the number of sources and targets is small the nodes visit probability is just an approximation of this quantity~\cite{alex-asta}. Considering this limit, we show the distribution of $p$ in Figure~\ref{probability} in both UCM and ER networks with $10^5$ nodes. The number of sources is $15$ and the target density is $0.2$. The curves show a power law behavior except for the presence of two peaks, representing the visit probability of sources and targets.
The peak for large values of $p_i$ is the consequence of the  visit of sources, and appears in correspondence of $p_i = p_{source}=1/N_S, \;\ \forall i \in \mathbf{S}$. The other peak is due to targets and occurs for $p_i = p_{targets}=1/N_T, \;\ \forall i \in \mathbf{T}$~\footnote{Both sources and targets are randomly selected, so we can assume that in average they will be visited $N_T$ times if they are sources or $N_S$ times if they are targets. Considering this and applying Eq.~\ref{eq:pi} we get $p_i=\rfrac{\sum \limits_{j=1}^{N_S N_T} \delta _{i,j}}{N_S N_T}=\frac{N_T}{N_SN_T}=\frac{1}{N_S}$ and $p_i=\frac{N_S}{N_SN_T}=\frac{1}{N_T}$, respectively}.

\section{Damage Detection}
\label{damage_det}

In order to introduce damage in the network, we consider that $N_D$ nodes are not functional, i.e. a fraction $\rho_D = N_D/N$, of nodes and all their links are removed from the network $\mathcal{G}$. We define the damaged network $\mathcal{G}_{D}(N_{D},M_{D})$ where the subscript $D$ denotes damage. 
Damaged nodes are selected either randomly or according to a degree based strategy in which nodes are removed according to their position in the degree ranking (hubs first or leafs first). Although probes target nodes can be damaged,  we assume that no sources are damaged. In these settings we aim at inferring the damage using shortest path sampling by looking at the number of nodes discovered before and after the damage occurs.

Shortest path probes are sent between each pair of source-target nodes. The sampled network after the damage $\mathcal{G}^*_D$  differs in general from the sampled view $\mathcal{G^*}$ of the original undamaged network because of the changed topology due to the missing nodes.
To quantify the damage we introduce the quantity
\be
\label{measure}
M=1-\frac{N^*_D}{N^*}
\ee
where $N^*$ and $N^*_D$ are the number of discovered nodes in the undamaged and damaged network respectively. 
If no damage occurs, the number of nodes discovered in $\mathcal{G}_D$ is similar to the one discovered in $\mathcal{G}$ so that $N^* \simeq N^*_D$ and $M \simeq 0$~\footnote{ The number $N^*$ may differ from $N^*_D$ also when no damage is present because of the not deterministic behavior of the shortest path algorithm. As we wrote in section~\ref{sampling} we use the RPS probing strategy, that return randomly one of the possibly equivalent shortest path hence providing different view of the same network}. If less nodes are seen in $\mathcal{G}_D$ than in $\mathcal{G}$ then $M > 0$, with $M=1$ representing the extreme case in which no nodes are discovered in the damaged network. Interestingly, the quantity $M$ can assume also negative values. Indeed, it is possible to see more nodes in $\mathcal{G}_D$ respect to $\mathcal{G}$. Although this case may sound counterintuitive at first, a closer look to the effect in the topology induced by removing nodes, clearly explain its meaning. Indeed, by removing some central nodes in the network (in the next section we discuss this point in details) the length of the shortest paths might increase on average as well as the number of discovered nodes.
\subsection{Numerical simulations}
\begin{figure}
\begin{centering}
\includegraphics[width=0.4\textwidth]{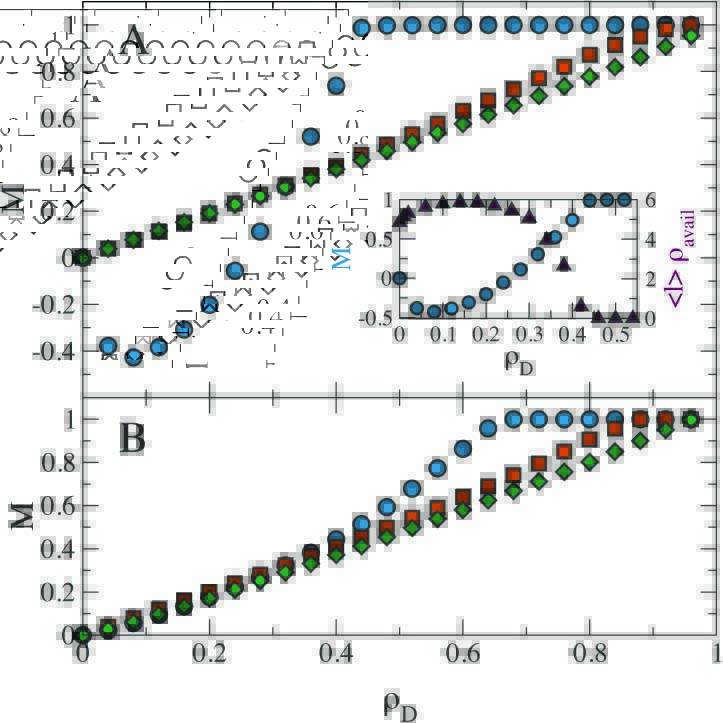}
\par\end{centering}
\caption{\label{01_damage_tot} {\small The behavior of $M$ is shown for three different damage strategies: random (red squares), small degree nodes first (green diamonds) hubs first (blue circles). (A) UCM scale-free graph. Inset: the behavior of $M$ when high degree nodes are removed first is compared to the quantity $\langle l \rangle \rho_{av}$ as a function of $\rho_D$. (B) ER random graph. For UCM network the minimum indicates the enhanced discovery given by the lack of hubs. Each plot is the median among $10$ independent assignment of sources and targets.}}
\end{figure}

We measure  the quantity $M$ in damaged homogenous and uncorrelated heterogenous networks~\cite{Newmanlibro, Alexlibro,Caldarellilibro,barabasi02}, generated through the ER and the UCM algorithm, respectively. The network size is fixed at $N=10^5$ nodes, and the number of sources and target density are  $N_{S}=15$, and $\rho_T=0.2$, respectively.
The average degree is $\bar{k}=8$ for both the topologies. The exponent $\gamma$ for UCM is $2.5$. As mentioned above, sources and targets are randomly selected.  We consider different damage strategies in which the removed nodes are selected at random or considering their degree. We further divide the latter strategy considering two cases in which nodes are removed in increasing (hubs first) or decreasing order of degree.

Figure~\ref{01_damage_tot} shows $M$ as a function of $\rho_D$ for the three different attack mechanisms, for the two different types of network. 
The top panel presents data for UCM network. The random nodes removal strategy gives the same qualitative behavior as the one in which the small degree nodes are attacked first. This is not surprising, since the probability that a randomly selected node has small degree is extremely high due to the power-law degree distribution of the network. The two strategies select, on average, the same category of nodes. A big difference can be noted when nodes are attacked in decreasing order of degree (high degree nodes, hubs, are attacked first). For small values of $\rho_D$ the value of $M$ assumes negative values, meaning that more nodes are discovered in the damaged network than in the undamaged one. Indeed, hubs act as shortcuts for network connectivity.  Their failure causes the rerouting of probes toward lower degree nodes and the consequent growth of the average length of the shortest paths $\langle l \rangle$.  As $\rho_D$ increases, this trend is contrasted by the progressive fragmentation of the network in many disconnected components.

In order to estimate how much the network has been fragmented by the damaging process we define the quantity $\rho_{av}$ as the average of the ratio between the nodes of the components in which each source is located, and the total number of nodes $N$ of the undamaged network. After a certain amount of nodes are removed, the graph undergoes disconnection and more than one component  appears. At this point a shortest path probe can reach only the nodes belonging to the component where the source is located. Components with no sources will be no longer accessible. $\rho_{av}$ is a decreasing function of $\rho_D$ assuming the value $1$ when there is no damage, whereas it becomes $\rho_{av} = N_S/N$ in the limit of $\rho_D \simeq 1$, when only the sources survive and each of them constitutes one component.
Neither $\langle l \rangle$ or $\rho_{av}$ alone explains the presence of the minimum of the quantity $M$ in the plots. Instead the product of the two $\langle l \rangle \rho_{av}$ does: it represents the average number of nodes discovered by each shortest path probe rescaled by the number of nodes effectively available to be discovered. The relation of this quantity with the minimum for $M$ is shown in the inset of Figure~\ref{01_damage_tot}A.
The argument above is confirmed by the behavior of $M$ in ER graphs. Here, 
removing the nodes with higher degree has a much smaller impact on the topology, and consequently there is no increase in the amount of nodes discovered in the damaged graph $\mathcal{G}_D$ with respect to the original one $\mathcal{G}$.
The plot of $M$ for hubs removal in the ER network does not show a minimum and substantially the damage detection works similarly for all damage strategies. 

\section{Single node damage detection}
\label{damage_sin}

While the measure $M$ quantifies the damage at the global level, it does not provide any information about specific nodes of the network. In this section we address the damage of individual nodes by assuming that the information gathered during the exploration of the undamaged network constitutes the null hypothesis of our measure, namely that no one of the nodes is damaged. We start by monitoring the network $\mathcal{G}$ assuming that it is not damaged. Every time we send a shortest path probe we obtain a better approximation of the sampled network $\mathcal{G}^*$ with increasing number of discovered nodes $N^*$. At the same time we collect information about how many times a probe passes through a node $i$ resulting on visit probability $p_i$ defined in Eq.~\ref{eq:pi}. The network is then damaged according to one of the strategies discussed above, and sampled via shortest path probes. By definition, any node that is discovered during the sampling is not damaged. However, the situation is less clear for nodes that have not been discovered. Indeed, the reason why a node is not seen can be either that it is actually damaged or that the sampling has missed it because the damage has altered the shortest path routing. 

\begin{figure}
\begin{centering}
\includegraphics[width=0.4\textwidth]{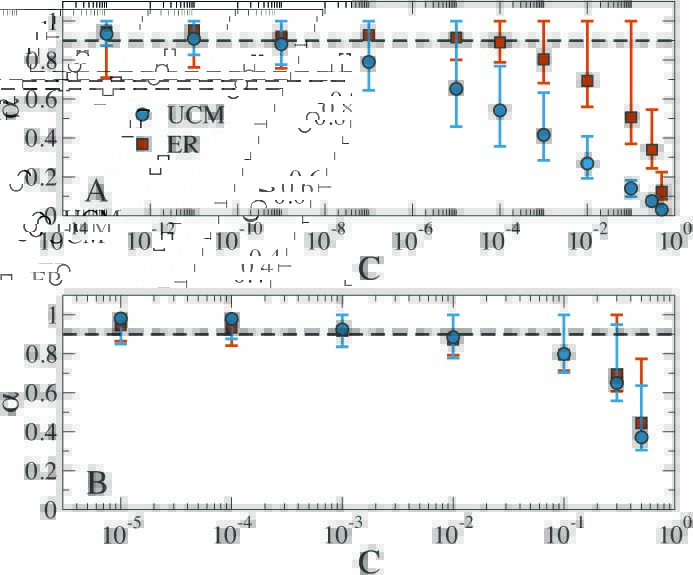}
\par\end{centering}
\caption{\label{precision_vs_C} {\small Precision $\alpha$ as a function of $C$ for (A) hubs removal with fraction of removed nodes $\rho_D = 0.001$ and (B) random nodes removal with fraction of removed nodes $\rho_D = 0.01$. The black dashed line indicate $\alpha=0.9$. Dots represent the average over 100 independent simulations and error bars illustrate the 95\% confidence interval.
}}
\label{fig:precision}
\end{figure}

In order to infer the state of undiscovered nodes we use a p-value test~\cite{Hanskibook1} applied to the visit probability $p_i$. More precisely, we calculate the probability $(1-p_i)^{\tau}$ of not seeing  the node $i$ after a number $\tau$ of shortest path probes. $\tau$ can assume any integer value from $1$ to $N_S \times N_T$. 
The p-value test consist in imposing the equality between this quantity and an arbitrary confidence level $C$:
\be
\label{p-value}
C = (1-p_i)^{\tau_i}
\ee
Note that after imposing the equality, $\tau_i$ has the index $i$ as for $p_i$. This is because $\tau_i$ is different for each node.
By taking the logarithm on both side of the equation we obtain:
\be
\label{tau}
\tau_i = \frac{\ln C}{\ln(1-p_i)}
\ee
If the node $i$ has not been seen at least once before $\tau_{i}$ probes have been sent then we can state that $i$ is damaged with statistical confidence $C$.
Here we are assuming that the visit probability of nodes does not change after the damage. This holds when the damage is a relatively small perturbation and does not change the connectivity of the network or its dynamical properties~\cite{havlin-book,Alexlibro,PhysRevE.88.062812}.
After the $p_i$ values have been determined for all the nodes, the value of $C$ tunes the number of probes to be sent before declaring a node damaged. If $C$ is selected to be large, nodes will be considered as damaged much earlier but with a small statistical confidence, leading to a large number of false positive ($FP$) detections. Conversely, if $C$ is set to be small, more probes are needed to state if a node is damaged or not. The accuracy improves and the final response will eventually return only actually damaged nodes, the true positive damaged nodes ($TP$). On the other hand the number of probes needed to reach this level of statistical confidence will be much higher resulting in a longer sampling process. 

\begin{figure}
\begin{centering}
\includegraphics[width=0.4\textwidth]{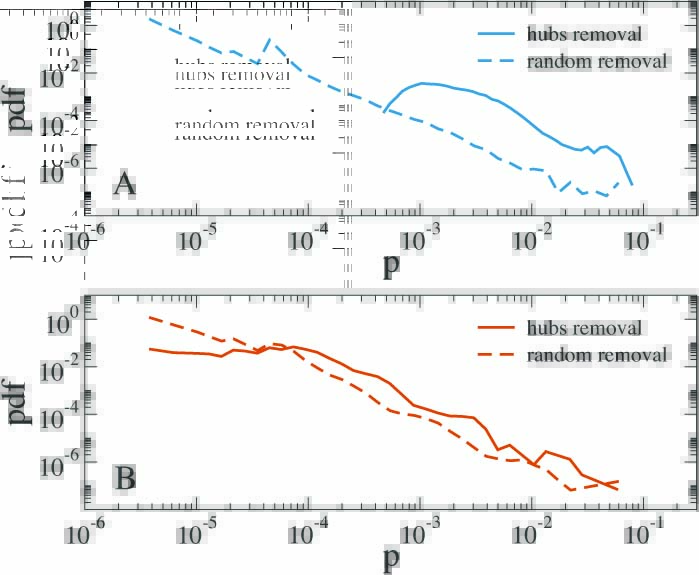}
\par\end{centering}
\caption{\label{fig:pidamaged} {\small Probability distribution function (pdf) of nodes visit probability in the undamaged UCM (A) and ER (B) networks restricted to nodes that will be later damaged with two different strategies.Curves are the average over 100 independent simulations.}}
\end{figure}

The value of $C$ is an input of the method and it can be chosen by opportunely tuning the trade-off between a poor but fast sampling that produces high number of $FP$s and an accurate but slow sampling that generates more $TP$s. We evaluate the performances of the damage detection strategy measuring its precision and recall. In particular, the precision, $\alpha$, is defined as
\be
\alpha=\frac{TP}{TP+FP}.
\ee
The recall, $r$, is instead
\be
r=\frac{TP}{TP+FN},
\ee
where $FN$ indicates the number of false negatives, i.e. nodes damaged but not detected. In a given network precision and recall are functions of the parameter $C$. In Figure~\ref{fig:precision} we plot $\alpha$ for different values of $C$ in both UCM and ER networks. In the top panel we remove top degree nodes and set $\rho_D=10^{-3}$. As expected $\alpha$ \begin{figure}
\begin{centering}
\includegraphics[width=0.4\textwidth]{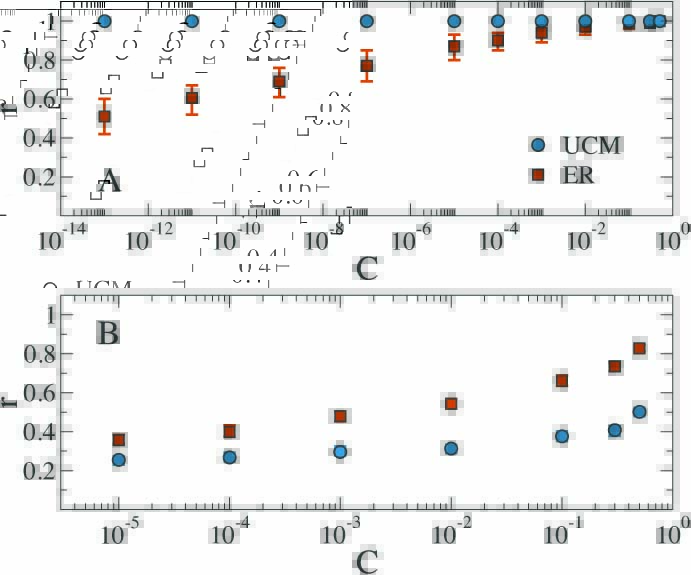}
\par\end{centering}
\caption{\label{recall_vs_C} {\small Recall $r$ as a function of $C$ for (A) hubs removal with fraction of removed nodes $\rho_D = 0.001$ and (B) random nodes removal with fraction of removed nodes $\rho_D = 0.01$. Dots represent the average over 100 independent simulations and error bars illustrate the 95\% confidence interval.
}}
\label{fig:recall}
\end{figure}
increases as $C$ decreases. Interestingly, in the case of UCM topologies the increase is slower. Indeed, we can notice that $\alpha$ reaches an arbitrary level of $90\%$ (dashed line) for $C=10^{-10}$ while in the case of ER networks the same level is reached for $C=10^{-5}$. The extremely low values of C, above all in the case of the UCM network, is justified by the presence of the logarithm function at the numerator of Eq.~\ref{tau}. Considering the big absolute value of the denominator for big $p_i$, very low values of $C$ are required to have values of $\tau_i$ that possibly ranges from 0 to $N_S \times N_T$. The quite different value of $C$ in the two networks for top degree nodes removal can be explained considering the distribution of $p_i$ of the damaged nodes in case of hubs removal as shown in Figure~\ref{fig:pidamaged}. In the UCM network top degree nodes have much higher visit probability respect to the rest of the nodes. In the ER network, instead, the role of high degree nodes is not so determinant and the visit probability of high degree nodes is almost indistinguishable from the one of random nodes. For the same value of $C$ in Eq.~\ref{tau} the difference in $p$ between UCM and ER translates in smaller values of $\tau_i$ for the UCM network. This leads to higher number of FPs and hence smaller precision. In Figure~\ref{fig:precision}B we show the same curve for the random damaging strategy setting $\rho_D=10^{-2}$. In this case the behavior of $\alpha$ in the two topologies is very similar. Such behavior is due to the $p$ distributions that in both networks span the entire range of possible visit probabilities irrespective of the topology reproducing the same behavior shown in Figure~\ref{fig:pi}.\\

In Figure~\ref{fig:recall} we study the recall $r$ as a function of $C$. In this case we can see that $r$ reaches 1 for all the values of $C$ investigated when damaging hubs in an UCM network. All damaged nodes are detected during the sampling. This is consequence of the very large visit probabilities for high degree nodes in the UCM network. Indeed, large $p_i$s combined with $C$ via Eq.~\ref{tau}, produce small values of $\tau_i$, allowing all removed nodes to be promptly declared damaged. The downside of this effect is the lack of precision for large values of $C$ (see Figure~\ref{precision_vs_C}). In the ER network, instead, the presence of low visit probability nodes among the ones in the top degree ranks makes their discovery more lengthy. Indeed, in this case larger values of $C$ are required to produce $\tau_i$s small enough to allow the algorithm to declare the nodes damaged.  It is crucial to stress that $\tau_{max}=N_S\times N_T$. Any node that for a given $C$ is characterized by $\tau_i > \tau_{max}$ will not be evaluated by the algorithm. 
In the bottom panel we see the recall for random nodes removal in both topologies. Here the behavior of $r$ is similar for both UCM and ER networks as consequence of the similar visit probability distributions.

\begin{figure}
\begin{centering}
\includegraphics[width=0.4\textwidth]{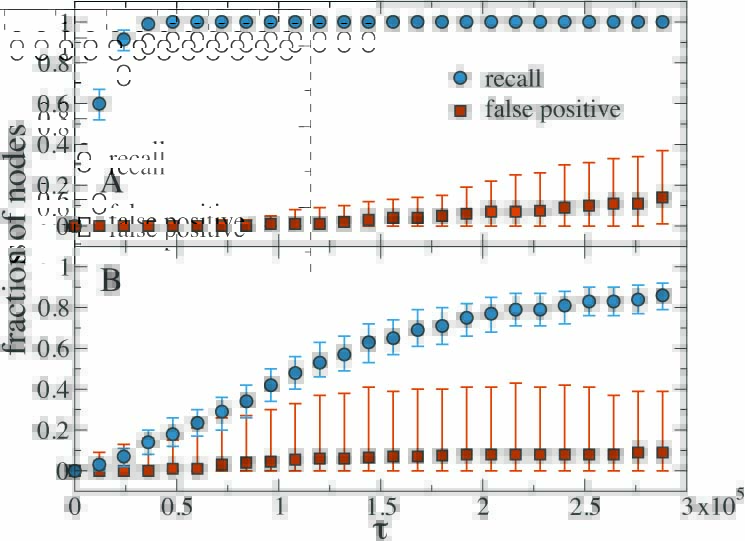}
\par\end{centering}
\caption{\label{tp_fp_big} {\small 
Recall $r$ (blu circles) and normalized number of false positive $fp$ (orange squares) for high degree nodes removal as a function of number of probes in UCM (A) and ER (B) networks. The fraction of removed nodes is $\rho_D = 0.001$. The values of confidence level $C$ are $10^{-10}$ and $10^{-5}$ for UCM and ER networks respectively. Points are the median among $100$ realization with independent choice of sources and target. Error bars illustrate the $95\%$ confidence interval.
}}
\end{figure}

\subsection{Numerical simulations in synthetic networks}
\label{det_damage_sin}
We apply the statistical criterion developed in the previous section to the two types of synthetic networks, UCM and ER, with $10^5$ nodes and two damaging strategies, high degree  and random nodes removal. We send shortest path probes from $15$ sources to a number of targets equal to a fraction $\rho_T = 0.2$  of total nodes. In order to compare the results of this part of the study for different topologies and damage strategies we arbitrarily fixed the $C$ value to the one correspondent to a precision of $\alpha=0.9$ in each system. \\
\begin{figure}
\begin{centering}
\includegraphics[width=0.4\textwidth]{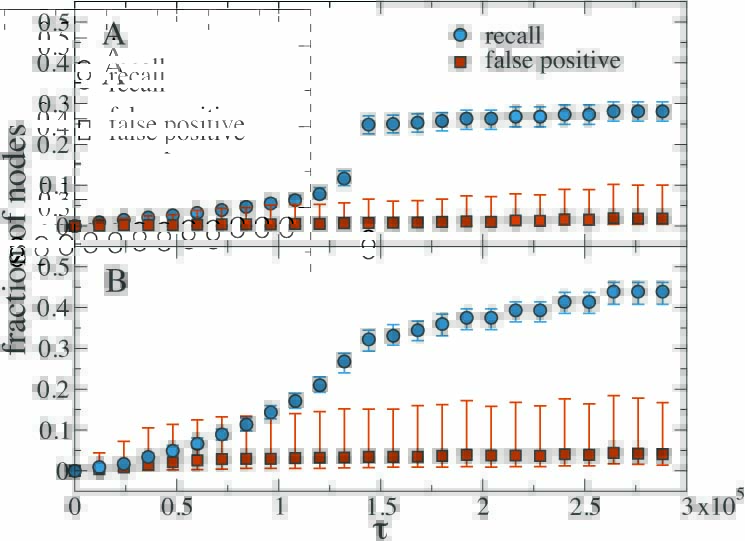}
\par\end{centering}
\caption{\label{tp_fp_random} {\small 
Recall $r$ (blu circles) and normalized number of false positive $fp$ (orange squares) for random nodes removal as a function of number of probes in UCM (A) and ER (B) networks. The fraction of removed nodes is $\rho_D = 0.01$.  The value of confidence level $C$ is $10^{-3}$ for both the UCM and ER network. Points are the median among $100$ realization with independent choice of sources and target. Error bars illustrate the $95\%$ confidence interval.}}
\end{figure}
Let us first consider UCM networks subject to the removal of the top hundred nodes ranked according to the  degree ($\rho_D = 10^{-3}$). In Figure~\ref{tp_fp_big} we plot the recall, $r$, and the normalized number of FPs, $fp=FP/(TP+FN)$ as a function of $\tau$. Interestingly, the recall reaches $1$ quickly. The absence of the hubs is promptly detected by the method. In Figure~\ref{tp_fp_random} we show the behavior of the same quantities in the case of random removal of nodes considering $\rho_D=10^{-2}$. In this case the recall increase slowly while $fp$ remains constant after an initial increase. An interesting feature of the $r$ curve is the presence of a jump. 
This jump is the consequence of the peak in the distribution of $p_i$ that is mapped into $\tau_i$ via Eq.~\ref{tau}. It occurs at the value of $\tau$ corresponding to $p_{targets}=1/N_T$ and is caused by the enhanced visit probability of target nodes.
Since targets are assigned randomly, in UCM networks they are most probably small degree nodes that are visited just if they are set to be targets. This imply that a specific number of probes equal to
\be
\tau_{targets} = \frac{\ln C}{\ln(1-p_{targets})}= \frac{\ln C}{\ln(1-1/N_T)}
\ee
is necessary to be able calling targets as damaged or not. Since targets are $20\%$ of the nodes, once $\tau_{targets}$ is reached a conspicuous amount of nodes is characterized by the p-value test. It is worth noting that only the number of $TP$s increases in correspondence with the jump, and $fp$ do not exhibit any discontinuity. This means that we have a better view of the damage without affecting the accuracy.\\
Let us now consider ER networks subject to removal of nodes in decreasing order of degree. In Figure~\ref{tp_fp_big}B we plot $r$ and $fp$ as a function of $\tau$. As for the case of UCM network, the recall increases, even if slower, and reaches the maximum values at $0.9$. Similar behavior for both the topologies is observed in the case of random removal of nodes (see Figure~\ref{tp_fp_random}).

\begin{figure}
\begin{centering}
\includegraphics[width=0.4\textwidth]{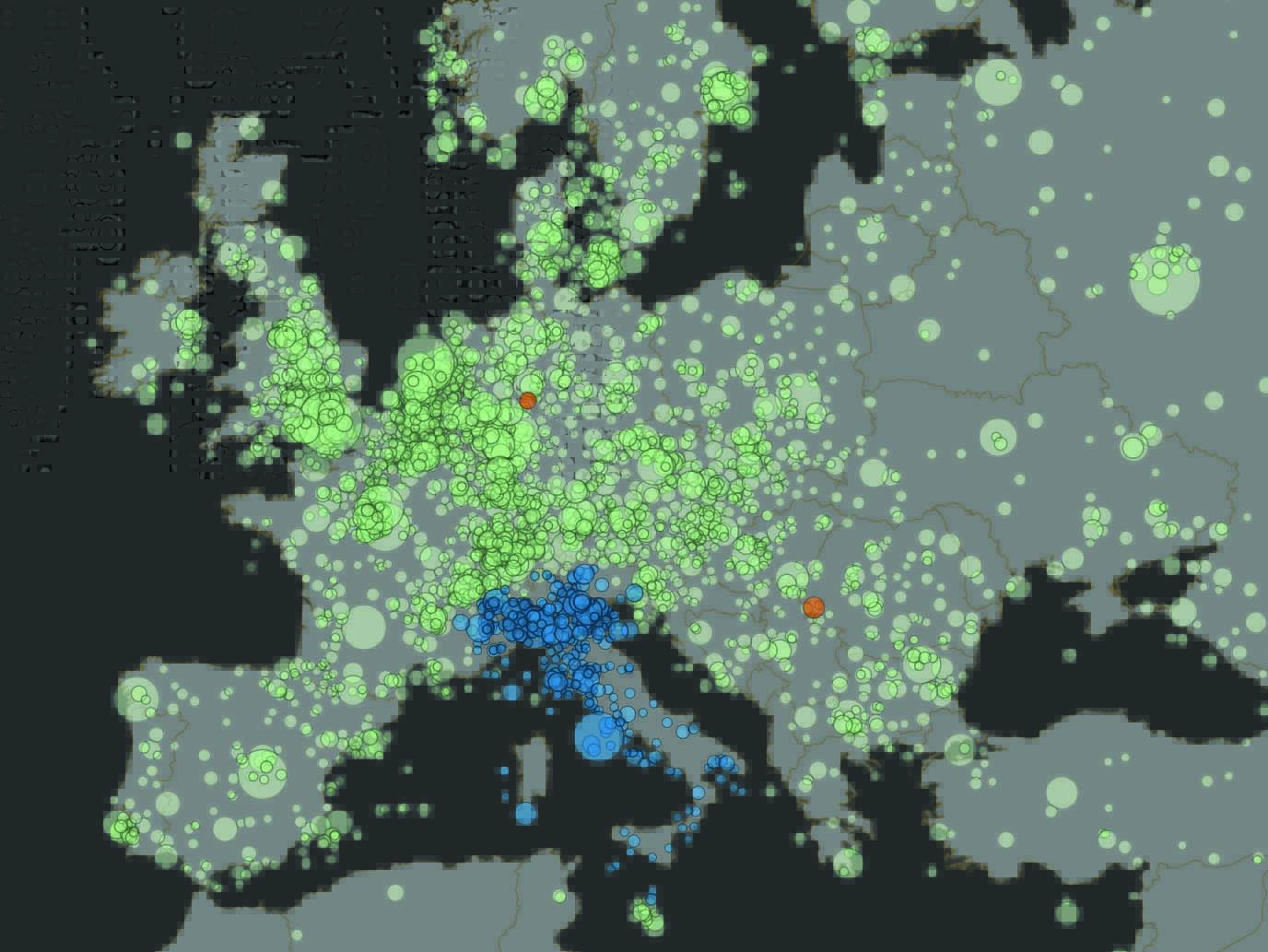}
\par\end{centering}
\caption{\label{italy} {\small Europe map showing the detected Italian damaged nodes in the real AS network. Each circle represent one AS and its size is proportional to the degree. Green circles are the working nodes, blue ones are the TPs while orange ones are the FPs. Nodes located at sea are effect of finite accuracy of geographical coordinates provided.}}
\end{figure}

\section{Numerical damage detection in geolocalized networks}
\label{internet_damage}
In this section we consider a sample of the real Internet topology network at the level of ASs where each node is an autonomous system of known geographical location~\cite{vaz-02,PsV04,caida}, and links represent the physical connections among them. Topologies are available for downloads in the DIMES project webpage~\cite{dimes}. We focus on the largest connected component of ASs that is made by $32852$ nodes.

We test damage detection in two relevant classes of realistic attacks: all nodes in the same country are attacked, and all nodes inside a radius $\xi$ with epicenter $E$ are attacked. Both of these strategies are geography-based but they provide different scenarios. The first represents a deliberate shut down, for instance as allegedly happened in several countries during the Arab spring~\cite{Dainotti_2011}. The second one is referred to localized event such as blackouts, earthquakes, or others catastrophic events~\cite{TechRepVirginiaTech2011}.
Also in this case we fix the number of sources $N_S = 15$ and the density of targets $\rho_T = 0.2$.
According to one of the two geographic based strategies we remove $N_D$ nodes from the original AS network.
The main difference between this case and those discussed in the previous sections is that the networks here have geographical attributes. The measure of damage detection should then be able to return not only the entity of the damage as a whole, but also tell us where the damage is localized.
\begin{figure}
\begin{centering}
\includegraphics[width=0.4\textwidth]{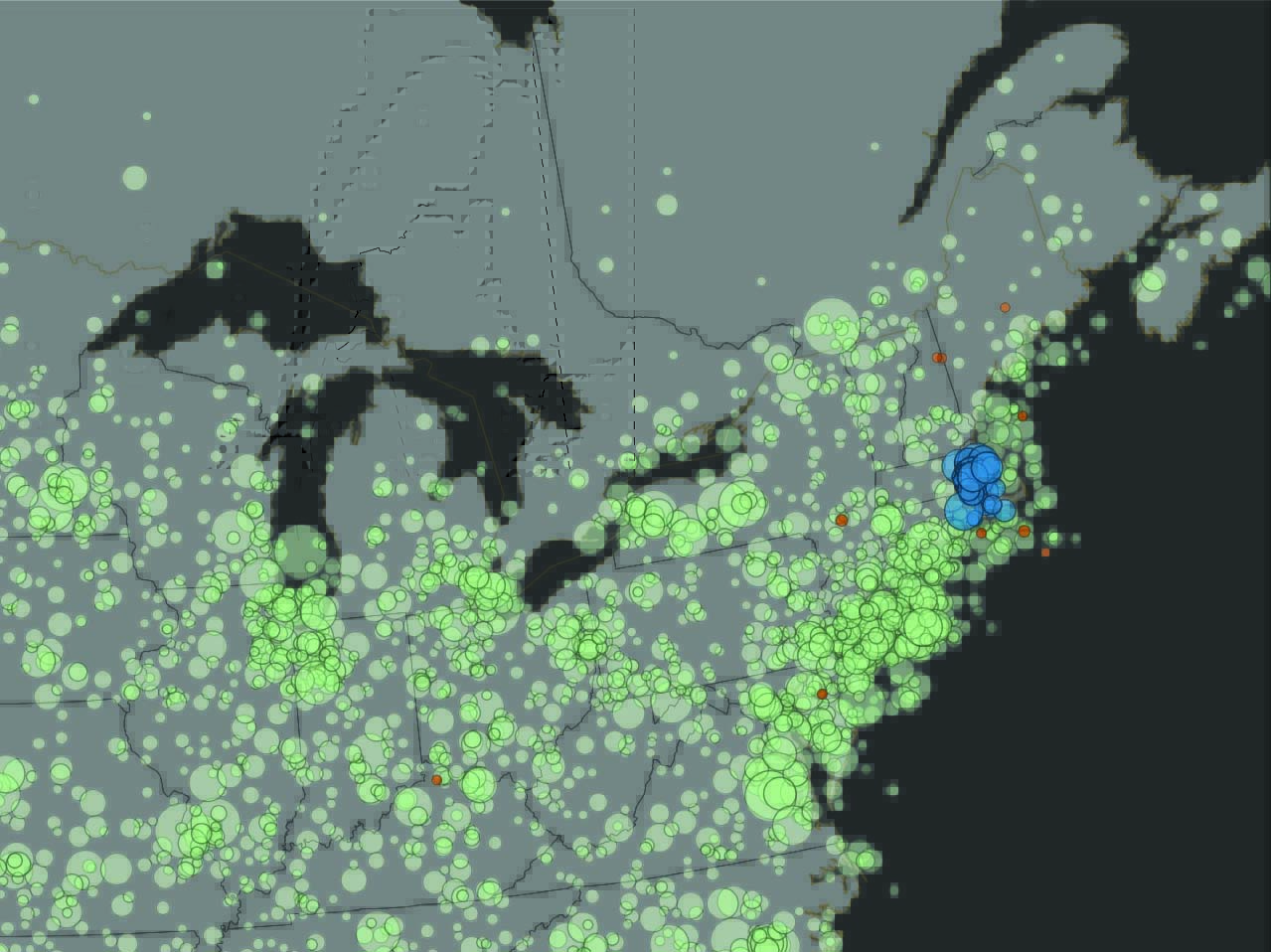}
\par\end{centering}
\caption{\label{boston} {\small Map of part of the United States east coast showing the outcome after damaging nodes around the city of Boston within a radius of $50$ km in the real AS network. Each circle represent one AS and its size is proportional to the degree. Green circles are the working nodes, blue ones are the TPs while orange ones are the FPs.}}
\end{figure}

\begin{figure}
\begin{centering}
\includegraphics[width=0.4\textwidth]{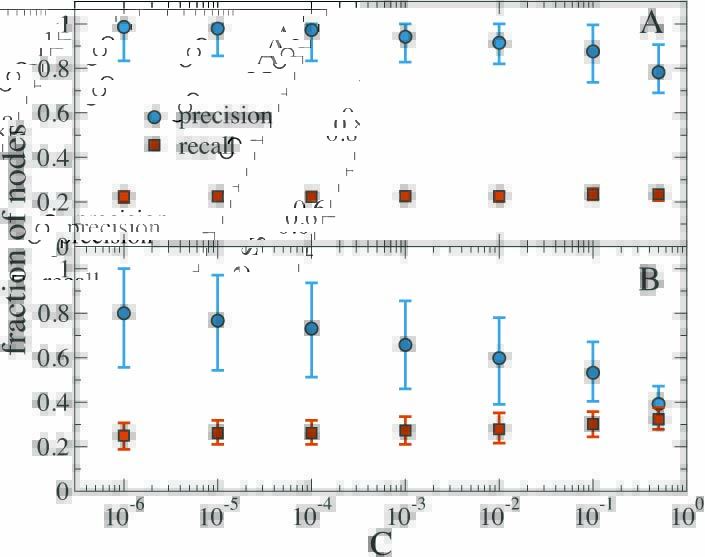}
\par\end{centering}
\caption{\label{geo_precision_recall} {\small Precision $\alpha$ (blu circles) and recall $r$ (orange squares) as a function of $C$ for Italian nodes removal (A) and damaging of nodes around the city of Boston within a radius of $50$ km (B) in the real AS network.
}}
\end{figure}

We use the same method already discussed for synthetic networks.
As an example of entire country switch off we decided to damage all the AS nodes in Italy. This translates in removing $N_D = 1246$ nodes, equal to a fraction $\rho_D = 0.038$ of total nodes. Figure~\ref{italy} shows the outcome of our analysis. We want to stress that the algorithm does not have any a priori information about the location of the damage. Despite that, the method clearly returns Italy as affected country. Few other nodes are wrongly classified as damaged. The reason for the presence of FPs can be just statistical fluctuations or, more interesting, that some FP nodes turn out to be strongly linked to the Italian TPs, so that the deletion of the latter prevents them to be visited.
For the second type of geographical damage we decide to switch off all the ASs within a radius of $50$ km around the city of Boston, MA, in the USA. This corresponds to $N_D = 176$ and $\rho_D=0.0054$. Also in this case the method is able to detect correct location of the damage as shown in Figure~\ref{boston}. 
In both cases of geographical damaging the recall is almost constant and close to the value 0.2 as shown in Figure~\ref{geo_precision_recall}. Indeed, the algorithm is detecting almost only the fraction of damaged nodes that are also target. Because of the homogeneous distribution of target nodes, this fraction corresponds to $\rho_T = 0.2$. The choice for the statistical confidence affects more the measures related to local damaging (Figure~\ref{geo_precision_recall}B). Here, for big values of $C$ the precision drops while the recall slightly increases. This means that a less strict choice of $C$ allows the discovery of more nodes. However the FPs grow more than the TPs. So a little gain in recall is contrasted by big loss in precision.
As for the artificial networks, in the case of entire country damaging we choose $C$ to achieve a precision of $0.9$ ($C= 10^{-2}$). In the case of local damaging there is no value of $C$ that allow to reach such a precision.  For this reason and considering the diverse nature of the two strategies we fix the arbitrary value to $\alpha = 0.75$ ($C= 10^{-5}$).
Despite the recall never exceeds 0.3 in both damage detections this is a good result considering the small number of damaged nodes, the completely random displacement of source and target nodes and the lack of any ad-hoc search strategy.

\section{Conclusions}
\label{conc}

In this paper we addressed the problem of damage detection in large-scale networks. We assessed the effectiveness of shortest path probing for damage detection in the case of incomplete network sampling. We considered different network topologies, damage strategies and defined basic metrics for the measurement of damage.  We provided a statistical criterion for the classification (damage/undamaged) of single nodes based on the p-value test. 
Although this criterion allows false positives, i.e nodes wrongly considered as damaged,  it is possible to fine tune the statistical confidence level in order to optimize the trade-off between precision and probing load in the system.  The numerical investigation according to this criterion allows the study of damage in partially sampled networks with tunable precision. In the case of real-world network such as the Internet AS graph, we damaged the network according to geographical features that simulate critical events on specific areas or deliberate shutdown of an entire country, as for political reasons.
Also in this cases, our methodology is able to identify the entity  of the damage and, more importantly, its location.

The method we have proposed can represent a first step towards a strategy for the continuous monitoring of large-scale, self-organizing networks. Possible variations of the shortest path sampling can be envisioned and combined with more elaborate  diffusive walkers strategies that optimizes network discovery. Furthermore, we have studied only the random displacement of sources and targets. Detection of damages could be improved by opportune choice of sources and targets or by different schedule of probes delivery. This points remain to be addressed in future works.

\section{Acknowledgments}
\label{akn}
The authors thank Bruno Gon\c calves for valuable discussions and the Dimes Project for providing Autonomous System topology free to download. This work was partially funded by the DTRA-1-0910039 award to A.V. The views and conclusions contained in this document are those of the authors and should not be interpreted as representing the official policies, either expressed or implied, of the Defense Threat Reduction Agency or the U.S. Government.




\end{document}